\documentclass[preprint,showpacs,preprintnumbers,amsmath,amssymb]{revtex4}

\usepackage{graphicx}
\usepackage{dcolumn}
\usepackage{bm}
\usepackage{tabularx}

\begin{document}

\title{Econophysics of Stock and Foreign Currency \\Exchange Markets}

\author{M. Ausloos$^{1}$}
\email{Marcel.Ausloos@ulg.ac.be}

\affiliation{ $^{1}$ SUPRATECS, Universit\'e de Li\`ege, B5 Sart-Tilman, B-4000 Li\`ege, Euroland}

\date{\today}


\begin{abstract}
Econophysics is a science in its infancy, born about ten years ago at this time of writing, at the crossing roads of  physics, mathematics, computing and of course economics and finance. It also covers human sciences, because all economics is ultimately driven by human decision. From this human factor, econophysics has no hope to achieve the status of an exact science, but it is interesting to discover what can be achieved, discovering potential limits and trying try to push further away these limits.  A few data analysis techniques are described with emphasis on the Detrended Fluctuation Analysis  ($DFA$) and the Zipf Analysis Technique ($ZAT$). Information about the original data aresketchy, but the data concerns mainly the foreign currency exchange market. The robustness of the $DFA$ technique is underlined. Additional remarks are given for suggesting further work. Models about  financial  value evolutions are recalled, again without  going into elaborate work discussing typical agent behaviors, but rather with hopefully sufficient information such that the basic ingredients can be memorized before reading some of the vast literature on price formation.  Crashes being spectacular phenomena  retain our attention and do so through data analysis and basic intuitive models.   A few statistical and microscopic models are outlined.

\end{abstract}

{\noindent \small \it An updated version of this article will be published as chap. 9 in a forthcoming book "Econophysics and Sociophysics: Trends and Perspectives", Eds. B.K. Chakrabarti, A. Chakraborti and A. Chatterjee, Wiley-VCH, Berlin.}


\maketitle

\section{Introduction}
Econophysics is a science in its infancy, born about ten years ago at this time of writing, at the crossing roads of  physics, mathematics, computing and of course economics and finance. It also covers human sciences, because all economics is ultimately driven by human decision. From this human factor, econophysics has no hope to achieve the status of an exact science, but it is interesting to discover what can be achieved, discovering potential limits and trying try to push further away these limits.

Numerous works have inspired physicists and guided them towards the  studies of financial markets. We could start with Bachelier \cite{Bach}  who introduced what shall perhaps remain the simplest and most  successful  model for price variations  but for the moment, it suffices to mention that it relies on Gaussian  statistics.  However, Bachelier could have learned from Pareto \cite{Pareto}  that power-laws are ubiquitous in nature. In fact, large price  variations in high frequency seems to be described by a crossover  between two power-laws, time correlations in the variance is also  decay as a power-law and so do numerous other quantities related to finance. Mandelbrot \cite{mandelbrot63} and Fama \cite{fama} helped move forward from these empirical evidences by proposing to describe price variations using a class of distributions studied by L\'evy \cite{Levy37}. Thereafter, these distributions were abruptly truncated by Mantegna and Stanley \cite{MaSt94}  and exponentially truncated by Koponen \cite{Koponen}  to recover a slow convergence towards a Gaussian distribution for low frequency data, as observed empirically.

From the previous collection of dates, going as far as 1897 for Pareto, one concludes that the field of research is not new, but the interest has been renewed recently as testified by the recent excellent works on  econophysics, starting with a conference book published by Anderson, Arrow and Pines \cite{AAP}  followed by the books of Bouchaud and Potters \cite{BP97}  and Mantegna and Stanley \cite{MaSt99}  Paul and Baschnagel \cite{PaBa99}  and Voit \cite{Voit}
These works show the variety of interests and successes of the approach of finance by physicists. From these works and the many research papers, we gather that there exists a temporary consensus on empirical data,   though there is no such unity in modeling. There is a profusion of  models indeed.
 
First recall   that a ``price'' can only go up or down, that is, is a one-dimensional quantity. Therefore, it can be seen as a point diffusing on a line. The factors driving this diffusive process are numerous, and their nature is still not fully elucidated. Hence, one can only make plausible assumptions and test afterwards their value by comparing the distribution of changes generated by the diffusion process with actual price evolution changes, price being also understood as  the value of some financial index or some exchange rate between currencies.

 A fundamental problem is the existence or not of long-range power-law correlations in economic systems as well as the presence of economic cycles. Indeed, traditional methods (like spectral methods) have corroborated that there is evidence that the Brownian motion idea is only approximately right \cite{mandelbrot63,fama}. Long-range power-law correlations   discovered in economic systems, particularily   in financial fluctuations \cite{MaSt94}  have been examined   through the so-called L\'evy statistics already mentioned by Stanley et al. \cite{MaSt94}  who have shown the existence of long-range power-law correlations in the Standard and Poor (S\&P500) index. A method based on wavelet analysis has also shown the emergence of hidden structures in the S\&P500 index \cite{ramsgy}. We have first  performed a Detrended Fluctuation Analysis ($DFA$) of the $USD/DEM$ ratio \cite{nvmafex} and of other  foreign currency exchanges and  have demonstrated the existence of successive sequences of economic activity having different statistical behaviors. They will be mentioned in the relevant section.

A  word of caution is at once necessary:  this review is by no means objective. It is strongly biased towards studies undergone by a few physicists in the  last few years, in order to give an overview of a few  results in the physics literature on only two topics :  foreign currency exchanges and  stock market indices. Even so, it is not possible to cover all papers published or put on arXives.  I do apologize for having missed many papers and ideas. Of course, since no full review of the different developments belonging to the physics literature is here possible, the more so about the purely economics literature. However on one hand, each of the research papers presented here has individually justified its sideline propositions which would be tiring to  reproduce. On the other hand,   there is   a huge amount of overlaps between the different  works summarized here below and  models developed by many authors. Trying to distinguish the main contributions separately is a  titan  work. Attempting to compare and discuss details is quasi impossible. Yet this review   of econophysics and data analysis techniques , restricted to stock markets and foreign exchange currency markets should only seen as to open a gate toward huge undiscovered fields. 

In the following section a few data analysis techniques will be described with emphasis on the Detrended Fluctuation Analysis  ($DFA$) and the Zipf Analysis Technique ($ZAT$). Information about the original data will be sketchy, but the data concerns mainly the foreign currency exchange market. The robustness of the  $DFA$ technique will be underlined   and additional remarks will be given for suggesting further work. In the next section models about  financial index value evolutions will be recalled, again without  going into elaborate work discussing typical agent behaviors, but rather with hopefully sufficient information such that the basic ingredients can be memorized before reading some of the vast literature on price formation.  Crashes being spectacular phenomena must retain our attention and do through data analysis and basic intuitive models.   A few ``more general''  microscopic  models will also be outlined.

\section{A few robust techniques}

\subsection{Detrended fluctuation analysis technique}

In the basic Detrended Fluctuation Analysis ($DFA$) technique one divides a time series   $y(t)$ of length $N$ into $N/\tau$ equal size nonoverlapping boxes \cite{DNA}. For smoothing out the data, the integration of the raw time series can often be performed first; in so doing one has to remember that such an integration shifts the exponent by one unit. The variable $t$ is discrete, evolves by a single unit at each time step between $t=1$ and $t=N$. No data point is supposed to be missing, i.e., breaks due to holidays and week-ends are disregarded. Nevertheless, the $\tau$ units are said to be $days$: often a week has only 5 days, and a year about 200 days. Let each box contain $\tau$ points and $N/\tau$ be an integer. The local trend in each $\tau$-size box is assumed to be linear, i.e. it is taken as $ z(t)$ = $a$  $t$ + $b$ .   In each $\tau$-size box one next calculates the root mean square deviation between $y(t)$ and $z(t)$. The detrended fluctuation function $F(\tau)$ is then calculated following
\begin{equation} 
{{F^2(\tau)}} = {1 \over \tau} {\sum_{t=k\tau+1}^{(k+1)\tau} {|y(t)-z(t)|}^2} , {\hskip 1cm} k=0,1,2, \cdots, ({N \over \tau} -1). 
\end{equation} 

Averaging $F^2(\tau)$ over all $N/\tau$ box sizes centered on time $\tau$ gives the fluctuations $\langle F^2(\tau) \rangle$ as a function of $\tau$.  The calculation is repeated for all possible different values of $\tau$. A power law behavior is expected as 
\begin{equation} 
\langle F^2(\tau) \rangle^{1/2} \sim \tau^{\alpha}. 
\end{equation} 
An exponent $\alpha \not= 1/2$ in a certain range of $\tau$ values implies the existence of long-range correlations in that time interval as in the fractional Brownian motion  \cite{west}. Such correlations are said to be ``persistent'' or ``antipersistent'' when they correspond to $\alpha >1/2$ and $\alpha <1/2$ respectively  in practice $Hu$ = $\alpha$.  A straight line can well fit the data between $log \tau$ = 1 and $log \tau$ = 2.6. This interval is called the $scaling~range$.    Outside the scaling range the error bars are larger often because of  so called finite size effects, and/or the lack of numerous data points. The $\alpha$ exponent is directly related to the Hurst exponent \cite{malamud} and the signal fractal dimension \cite{mandelbrot63}.

Most of the time, for the real or virtual foreign exchange currency ($FEXC$) rates that we have examined \cite{nvmafex,physica1,malatin,ladek,ma13,ma14,ma15,ma18,ma19,ma20,ma24}, the scaling range is well defined. Sometimes it readily appears that the data contains two sets of points which can be fitted by straight lines. Usually that describing the ``large $\tau$'' data has a $0.50$ slope, Indicating correlationless fluctuations.   Crossovers from fractional to ordinary Brownian motion can be well observed. These crossovers suggest that correlated sequences have characteristic durations with well defined lower and upper scales \cite{nvmafex}.  The persistence is usually related to free market (and ``runaway'') conditions while the antipersistence develops due to strict political control allowing for a finite size bracket in which the $FEXC$ rates can fluctuate. The case $\alpha = 0.50$ is surely avoided by speculators.

Notice that to consider overlapping boxes might be useful when not many data points are available. However it was feared that extra insidious correlations would thereby be inserted. Nevertheless the analysis has shown that the value of $\alpha$ is rather insensitive to the way boxes are used.  A cubic trend, like 
$z(t)= c t^3+ d t^2 + e t+ f$, can be also considered  \cite{ndub}. The parameters $a$ to $f$ are similarly estimated through a best least-square fit of the data points in each box. Following the procedure described here above the value of the exponent $\alpha$ can be obtained. Again the difference is found to be very small. Extensions of  $DFA$ to higher moment components have also been investigated, tending toward some sort of multifractal analysis. Moreover the linear or cubic or other polynomial law fitting the trend can be  usefully replaced by some moving average, leading to some further insight into the Hurst exponent value.\cite{anna}

An interesting observation consists at looking for \textit{local} (more exactly 
\textit{temporal}) $\alpha$ values   \cite{nvmafex,malatin,ndub}. 

This   allows one to probe the existence or not of {\it locally correlated or decorrelated sequences}.     In order to observe the local correlations, a local observation box of a given finite size is constructed. Its size depends on the upper value of $\tau$ for which a reasonable power law exponent is found. It is chosen to be large enough in order to obtain a sufficiently large number of data points. The $local$ exponent $\alpha$ is then calculated for the data contained in that finite size box, as above. Thereafter the box is moved along the time axis by an arbitrary finite number of points, often depending on the intended strategy.  

The $local~\alpha~exponent$    seems rougher, and varies with time around the overall (mean) $\alpha$ value. The variation depends on the box size. Three typical $FEXC$ rate time dependences, i.e. $DEM/JPY$, $DEM/CHF$, and $DEM/DKK$ have been shown for various time intervals in \cite{ladek} and the $\alpha$ value indicated for the scaling range.  For example, the $DEM/JPY$ $local~\alpha$ is consistently above $0.50$  indicating  a persistent evolution. A positive fluctuation is likely to be followed by another positive one. The case of $DEM/CHF$ is typically Brownian with fluctuations around $0.50$. However, in 1994 some drift is observed toward a value $ca.$  $0.55$ while in 1997-1998 some drift is observed toward a value $ca.$ $0.45$. It is clear that some economic policy change occurred in 1994, and a drastic one at the end of 1998 in order to render the system more Brownianlike. The same is true for the $DEM/DKK$ where due to european economic policy the spread in this exchange rate was changed several times, leading from a Brownianlike situation to a nowadays antipersistent behavior, i.e. a positive fluctuation is followed by a negative one, and conversely, such that $local~\alpha$ is becoming pretty low.

Therefore this procedure interestingly leads to a $local$ measurement of the degree of long-range correlations. In other investigations,  \cite{nvmafex} it has been found on   the $GBL/DEM$ exchange rate data that the change in slope of $local~\alpha$ $vs.$ time corresponds to changes in the Bundesbank interest rate increase or decrease.  The 1985 Plaza agreement   had some influence in order to curb the run away $local~\alpha$ value from a persistent $0.60$ back to a more $0.50$ Brownian-like value. Thus such $FEXC$ rate behaviors indicate a measure of information, an entropy variation indicating how (whatever) ``information
 is managed by the system. This seems to be an ``information'' to be taken into account when developing Hamiltonian or thermodynamic-like models.

At some time,\cite{ma13,ma14} it was also searched for correlations and anticorrelations in exchange rates of pre-Euro moneys with respect to currencies as $CHF$, $DKK$, $JPY$ and $USD$ in order to understand the $EUR$ behavior. The power law behavior describing the rms. deviation of the fluctuations as a function of time, whence the Hurst (or $\alpha$)   time-dependent exponent was obtained.   We compared the time dependent $\alpha$ exponent of the  $DFA$ as in a correlation matrix for estimating respective influences.  In doing a similar analysis, it was shown that the fluctuations of the $GBP$ and $EUR$ with respect to the major currencies were very similar, indirectly indicating that the $GBP$ was effectively tied to $EUR$. Same for studies pertaining to latin american currencies \cite{malatin}.
 
Since such temporal correlations can be sorted out, a strategy for profit making can be developed. It is easily observed whether there is persistence or antipersistence in some exchange rate, - according to the temporal value of $\alpha$. Thus some probability on the next fluctuation, to be positive or negative  can be made. Therefore a buy or sell decision can be taken.  In so doing and taking the example of ($DEM/BEF$), we performed some virtual game and implemented the most simple strategy 
\cite{dublin2}.
   
Investment strategies should look for $Hu$ values over different time interval windows which are continuously shifted. In this respect connection to multifractal analysis.  The technique consists in calculating the so-called ``$q$th order height-height correlation function''  
\begin{equation} 
c_q(\tau) = {\langle {|y(t)-y(t') |}^q \rangle}_{\tau} 
\end{equation} 
where only non-zero terms are considered in the average 
${\langle ... \rangle}_{\tau}$ taken over all couples $(t,t')$ such that 
$\tau = |t-t'|$. The correlation function {\it c({$\tau$})} is supposed 
to behave like 
\begin{equation} 
c_1(\tau) = {\langle {|y(t)-y(t')|} \rangle}_{\tau}  \sim {\tau}^{H_1}, 
\end{equation} 
where $H_1$ is the $Hu$ exponent. \cite{ref13,r1} This corresponds to obtaining a spectrum of moving averages indeed ~ \cite{mobavnvma}. Notice that a $DFA$ and a multifractal analysis cost much more CPU time than a moving average method due to multiple loops which are present in the $DFA$ and $c_q$ algorithms     \cite{nvmafex,ref13,mobavnvma,multiaffnv}.
 
Of course, subtracting some moving average background, instead of the usual linear trend, is of interest in order to implement a strategy based on different horizons. 

\subsection{Zipf analysis technique}

The same type of consideration for a strategy can be developed from the  Zipf analysis technique ($ZAT$) performed on stock market or $FEXC$ data. In the ($DFA$) technique  the sign of the fluctuations and their persistence, are taken into account, but it falls short of implementing some strategy from the {\it amplitude} of the fluctuations. The so-called  Zipf analysis \cite{zipf} originally introduced in the context of  natural languages can be performed by calculating the frequency of occurence $f$ of each word in a given text.   According to their frequency, a rank $R$ can be assigned to each word, with $R=1$ for the most frequent one. A power law 
\begin{equation} 
f~ \sim ~R^{-\zeta}  
\end{equation} 
is expected  \cite{zipf}. A  Zipf plot is simply a transformation of the cumulative distribution. However, it accentuates the upper tail of the  distribution, which makes it a useful to analyze this part of the  distribution.

A simple extension of the Zipf analysis is to consider $m$-letter words, i.e. the words strictly made of $m$ characters without considering the white spaces. The available number of different letters or characters $k$ in the alphabet should also be specified.   A power law in $f(R)$ is expected as well for correlated sequences \cite{z1,z2}. There is no theory at this time predicting the exponent as a function of ($m,k$). The technique has a rather weak value when only two characters and short words are considered  \cite{z1}.   An increase in the number of allowed characters for the alphabet allows one to consider  different size and signs of the fluctuations, i.e. huge, marginal and small (positive or negative) fluctuations can be considered.   After having decided on the number ($k$) of characters  of the alphabet, the signal can be transformed into a text and thereafter analysed with respect to the frequency of words of a given size ($m$) to be ranked accordingly. In our work, words of equal lengths were always considered 

The above procedure does not take into account the trend.  Some work has been done on the matter, but much is still to be performed  since the trend definition can be quite different for later strategy considerations.  The relevance of this remark should be emphasized:  indeed for a positive (or negative) trend over the time box which is investigated, a bias can occur between words. For a two character alphabet, e.g. $u$ and $d$, the frequency $f$ of $u$'s, i.e. $p_u$ can be larger (smaller) than the frequency of $d$'s, i.e. $p_d$.  Such a $bias$ can be taken into account with respect to the equal probability occurrence, e.g. $\epsilon$ = $p_u~-~0.5$ =$p_d~+~0.5$.  A new ranking procedure can be performed by defining  the ratio of the observed frequency of a word divided by the theoretical frequency of occurrence of a word, assuming independence of characters. E.g. if the word $uud$ occurs say $p_{uud}$ times, since the independence of characters would imply that the word would occur $p_u~p_u~p_d$ times, a relative frequency $f/f_0$ can be defined as $p_{uud}/(p_u~p_u~p_d)$. A new ranking can be made with quite different appearance.

A  $ZAT$ variant consists in ranking the words according to their relative frequency and relative (``normalized'') rank taking into account for the normalization the probability $f_M$ of the most often occurring word \cite{z2}. Indeed  for $m$ and $k$ large not all words do occur. Even though, e.g. for the ($m=6,k=2$) case there are 64 possible words, and the maximum rank is $R_M$ = 64, the frequency of the most often observed word is unknown.   Another extension has been recently proposed in which  time windows are used in order to sort better the relevant exponents and probability of occurrence \cite{bronlet1,bronlet2}.

In conclusion of this subsection, it is emphasized  that different strategies following the Zipf analysis technique can be implemented, according to $(m,k)$ values, how the trend is eliminated (or not) and how the ranks and frequencies of occurrence are defined.

\subsection{Other techniques for searching correlations in financial indices} 
 
Quite simultaneously, Laloux {\it et al.} \cite{laloux1,laloux2} and Plerou {\it et al.} \cite{plerou1,plerou2} analyzed the correlations between stocks traded on financial markets using the theory of random matrices. Laloux {\it et al.}   considered daily price changes for the 1991-1996 period (1309 day) of 406 of the companies forming the S\&P500 index while Plerou {\it et al.} analyzed price  returns over a 30 minute period of the 1,000 largest US companies for the 1994-1995 period (6448 points for each company). The correlation coefficient between two stocks $i$ and $j$ is defined by 
\begin{equation} 
\rho_{ij} \equiv \frac{<R_i R_j> - <R_i> <R_j>} {\sqrt{(<R_i^2> - <R_i>^2)(<R_j^2> - <R_j>^2)}} \label{eq:correlation coefficient} 
\end{equation} 
where $R_i$ is the price of company $i$ for Laloux {\it et al.} and the return of the price for Plerou {\it et al}. The  statistical average is a temporal average performed on all trading  days of the investigated periods. The cross-correlation matrix $C$ of elements $\rho_{ij}$ is measured at equal times.

Laloux {\it et al}  \cite{laloux1,laloux2}  focussed on the density $\rho_c (\lambda)$ of eigenvalues of $C$, defined by 
\begin{equation} 
\rho_c (\lambda) \equiv \frac{1}{N} \frac{d n (\lambda)}{d\lambda} 
\end{equation} 
where $n (\lambda)$ is the number of eigenvalues of $C$ less than  $\lambda$ and $N$ the number of companies (or equivalently the  dimension of $C$). The empirically determined $\rho_c (\lambda)$ is  compared with the eigenvalue density of a random matrix, that is, a  matrix whose components are random variables. Random matrix theory  (RMT) isolates universal features, which means that deviations from these features identifies system-specific properties. Laloux {\it et al.} found that 94\% of the total number of eigenvalues fall inside the prediction of RMT, which means that it is very difficult  to distinguish correlations from random changes in financial markets.

The most striking difference is that RMT predicts a maximum value for the largest eigenvalue which is much smaller than what is observed empirically. In particular, a very large eigenvalue is measured, representing the market itself. Plerou {\it et al.} \cite{plerou1,plerou2} observed similar properties in their empirical study. They measured the number of companies contributing to each eigenvector and found that a small number of companies (compared to 1,000) contribute to the eigenvectors associated to the largest and the smallest eigenvalues. For the largest eigenvalues these companies are correlated, while for the smallest eigenvalues, these are uncorrelated. The observation for the largest eigenvalues does not concern the largest one, which has an associated eigenvector representing the market and has an approximately equal contribution from each stock.

Mantegna \cite{mantegna99} analyzed the correlations between the 30 stocks used to calculate the Dow Jones industrial average and the correlations between the companies used to calculate the S\&P500 index, for the July 1989 to October 1995 time period in both cases. Only the companies which were present in the S\&P500 index during the whole period time were considered, which leaves 443 stocks. The correlation coefficient of the returns for all possible pairs of stocks was computed.  A metric distance between two stocks is defined by $d_{ij} = 1 - \rho_{ij}^2$. These distances are the elements of the distance matrix $D$.

Using the distance matrix $D$, Mantegna determined the topological arrangement which is present between the different stock. His study could also give empirical evidence about the existence and nature of common economic factors which drive the time evolution of stock prices, a problem of major interest. Mantegna determined the minimum spanning tree (MST) connecting the different indices, and thereafter, the subdominant ultrametric structure and heriarchical organization of the indices. In fact, the elements $d^{<}_{ij}$ of the ultrametric distance matrix $D^{<}$ are determined from the MST. $d^{<}_{ij}$ is the maximum Euclidian distance $d_{lk}$ detected by moving by single steps from $i$ to $j$ through the path connecting $i$ to $j$ in the MST. Studying the MST and hierarchical organization of the stocks defining the Dow Jones industrial average, Mantegna showed that the stocks can be divided in three groups. Carrying the same analysis for the stocks belonging to the S\&P500, Mantegna obtained a grouping of the stocks according to the industry they belong to. This suggests that stocks belonging to the same industry respond in a statistical way to the same economic factors.

Later, Bonanno {\it et al.} \cite{bonanno} extended the previous analysis to consider correlations between different (29 stock market indices, one from Africa, eight from America, nine from Asia, one from Oceania and ten from Europe. Their correlation coefficients $\rho_{ij}$ was calculated using the return of the indices instead of the individual stocks. They also modified slightly the definition of the distance, using $d_{ij} = \sqrt{2 (1 - \rho_{ij})}$. A hierarchical structure of indices was obtained, showing a distinct regional clustering. Clusters for North-America, South America and Europe emerge, while Asian indices are more diversified. Japanese and Indian  stock markets are pretty distant from the others. When the same analysis is performed over different periods of time, the clustering is still present, with a slowly evolving stucture of the ultrametric distance with time. The taxinomy is stable over a long period of time.

Two key parameters are the length $L^{<}$ and the proximity $P$. The length is $L^{<} \equiv \sum d^{<}_{i,j}$, where the sum runs over nearest neighbouring sites on the MST. It is a kind of average of first-neighbour distances over the whole MST. The proximity is
\begin{equation} 
P \equiv \frac{\sum_{i,j} |d_{i,j} - d^{<}_{i,j}|}{\sum_{i,j} d_{i,j}} 
\end{equation} 
where the sums run over all distinct $i,j$ pairs. The proximity characterizes the degree of resemblance of the subdominant ultrametric space to the Euclidian space. For long time period, $L^{<}\to 26.9$, while $L^{<} = 28\sqrt{2} \approx 39.6$ for sequences without correlations. Similarly, for long time period, $P  \to 0.091$, to compare with $P \to 0.025$ when the same data are shuffled to destroy any correlations. The effects of spurious cross-correlations make the previous time analysis relevant for time periods of the order of three months or longer for daily data.

Others  \cite{VBB}  analyzed the sign of daily changes of the NIKKEI, the DAX and the Dow Jones industrial average  for the Dow Jones only and  for the three time series together, reproducing the evolution of the market through the sign of the fluctuation, as if the latter was represented by an Ising spin. The authors studied the frequency of occurrence of triplets of spins for the January 1980 to December 1990 period, after having removed the bias due to a different probability of having a move upward or downward. They showed that the spin series generated by each index separately is comparable to a randomly generated series. However, they emphasized correlations in the spin series generated by the three indices showing that market places fluctuate in a cooperative fashion. Three upward moves in a row are more likely than expected from a series without correlations between successive changes, and similarly for three downward moves. This behaviour seems to be symmetrical with respect to ups and downs in the time period investigated. The strength of the correlations varies with time, with the difference in frequency of different patterns being neatly marked in the two year period preceding the 1987 crash. Also, during this period, the up-down symmetry is broken.

Finally let us mention the Recurrence Plot ($RP$) and Recurrence Quantification Analysis  ($RQA$) techniques for detecting e.g. critical regimes appearing in financial market indices \cite{AFMAIJMPC}. Those are graphical tools elaborated by Eckmann, Kamphorst and Ruelle in 1987, based on Phase Space Reconstruction \cite{rp} and extended by 
Zbilut and Webber \cite{crq} in  1992 $RP$ and $RQA$ techniques are usually intended to provide evidence for chaos, but can also be used for their goodness in working with non stationarity and noisy data \cite{crqnoise} and in detecting changes in data behavior, in particular in detecting breaks, like a phase transition \cite{lambertz},  and other dynamic properties of a time series \cite{rp}. It was indicated that they can be used for detecting bubble nucleation, grow and even indicate bursting. An analysis was made on two time series, NASDAQ and DAX, taken over a time span of 6 years including the known (NASDAQ)  crash of April 2000. It has been shown that  a bubble burst warning could be given, with some delay ($(m-1)d$ days) with respect to the beginning of the bubble, but  with enough time before the crash (3 months in the mentioned cases) \cite{AFMAIJMPC}.

\section{Statistical, phenomenological and ``microscopic'' models}

A  few functions used to fit empirical data should be mentioned. The  point is to try to determine which laws are the best approximation for which data. Some of these laws rely on plausible and fairly convincing arguments  A major step forward to quantify the discrepancies between real time  series and Gaussian statistics was made by Mandelbrot \cite{mandelbrot63} who studied cotton prices. In addition to being non-Gaussian, Mandelbrot noted that returns respect time scaling, that is, the distribution of returns for various time scales $\Delta t$ from 1 day up to one month have similar functional form. Motivated by this scaling and the fact that large events are far more probable than expected, he proposed that the statistics of  returns could be described by symmetrical L\'evy distributions; other laws can appear to better fit the data, and also rely on practical arguments, yet they appear as {\it ad hoc} modifications, like the truncated L\'evy flight. 

\subsection{ARCH, GARCH, EGARCH, IGARCH, FIGARCH models}

The acronym `ARCH' stands for autoregressive conditional heteroscedasticity, a process introduced by Engle \cite{engle}. In short, Engle  assumed  that the price at a given time is drawn from a probability distribution which depends on information about previous prices, what he referred to as a conditional probability distribution (cpd). Typically, in a ARCH($p$) process, the variance of the cpd $P (x_t| x_{t-1}, ..., x_{t-p})$ at time $t$ is given by
\begin{equation} 
\sigma_t^2 = \alpha_0 + \alpha_1 x^2_{t-1} + \alpha_2 x^2_{t-2} + ... \alpha_p x^2_{t-p} 
\end{equation}  
where $x_{t-1}$, $x_{t_2}$, ... are random variables chosen from sets of random variables with Gaussian distributions of zero mean and standard deviations $\sigma_{t-1}$, $\sigma_{t-2}$, ..., respectively. $\alpha_0$, $\alpha_1$, ..., $\alpha_p$ are control parameters. Locally (in time), $\sigma_t$ varies but on long time scale, the overal process is stationary for a wide range of values of the control parameters.
 
Bollerslev \cite{bollerslev} generalized the previous process by introducing Generalized ARCH or GARCH($p$,$q$) processes. He suggested to model the time evolution of the variance of the cpd $P (x_t| x_{t-1}, ...,  x_{t-p}, \sigma_{t-1}, ... \sigma_{t-q})$ at time $t$ with
\begin{equation} 
\sigma_t^2 = \alpha_0 + \alpha_1 x^2_{t-1} + ... \alpha_n x^2_{t-n} +  \beta_1 \sigma^2_{t-1} + ... + \beta_q \sigma^2_{t-q}, 
\end{equation} 
with an added set $\{ \beta_1, ..., \beta_q \}$ of $q$ control parameters on top of already $p$ control parameters. Using this process, it is now possible to generate time correlations in the variance. The previous processes have been extended, as in EGARCH processes \cite{nelson}, IGARCH processes \cite{engleboll86}, FIGARCH processes \cite{BBM}, among others.

\subsection{Distribution of returns}
 
Mantegna and Stanley \cite{MaSt95} analyzed the returns of the S\&P500 index from Jan. 1984 to Dec.  1989. They found that it is well described by a L\'evy stable symmetrical process of index $\alpha = 1.4$ for time intervals spanning from 1 to 1,000 minutes, except for the most rare events. Typically, a good agreement with the L\'evy distribution is observed when $m/\sigma \le 6$ and an approximately exponential fall-off from the stable distribution is observed when $m/\sigma \ge 6$, with the variance $\sigma^2 = 2.57.10^{-3}$. For $\Delta t = 1 minutes$, the kurtosis $\kappa = 43$. Their empirical study relies mainly on the scaling of the `probability of return to the origin' $P (R = 0)$ (not to be confused with the distribution of  returns) as a function of $\Delta t$, which is equal to \begin{equation} 
P (0) \equiv {\cal L}_{\alpha} (R_{\Delta t} = 0) = \frac{\Gamma  (1/\alpha)}{\pi \alpha (\gamma \Delta t)^{1/\alpha}}  \end{equation} 
if the process is L\'evy, where $\Gamma$ is the Gamma function. They obtain $\alpha = 1.40 \pm 0.05$. This value is roughly constant over the years, while the scale factor $\gamma$ (related to the vertical position of the distribution) fluctuates with burst of activity localized in specific months. Using a ARCH(1)  model with $\sigma^2$ and $\kappa$ constrained to their empirical values, Mantegna and Stanley \cite{MaSt98} obtained a scaling value close to 2 ($\sim 1.93$), in disagreement with $\alpha = 1.4$. For a GARCH (1,1) process with similar constraint and the choice $\beta_1 = 0.9$, In ref. \cite{MaSt95,MaSt98} they obtained a very good agreement for the distribution of returns when $\Delta t = 1 minute$, but they estimated that the scaling index $\alpha$ should be equal to $1.88$, in disagreement with the observed value.
 
Gopikrishnan {\it et al.} \cite{Gopi99} extended the previous study of the S\&P 500 to the 1984-1996 period for one-minute frequency records, to the 1962-1996 period for daily records and the 1926-1996 period for monthly records. In parallel, they also analyzed daily records of the NIKKEI index of the Tokyo stock exchange for the 1984-1997 period and daily records of the Hang-Seng index of the Hong Kong stock exchange for the 1980-1997 period. In another study, Gopikrishnan {\it et al.} \cite{Gopi98} analyzed the returns of individual stocks in the three major US stock markets, the NYSE, the AMEX and the NASDAQ for the January 1994 to December 1995 period. Earlier works by Lux \cite{lux} focussed on daily  returns of the German stocks making the DAX share index during the period 1988-1994.

To compare the behaviour of the different assets on different time scales, a normalized  return $g \equiv g_{\Delta t} (t)$ is defined, with 
\begin{equation}  
g \equiv \frac{R - <R>_T}{\sqrt{<R_T^2> - <R>_T^2}} 
\label{eq:normalized return} 
\end{equation} 
where the average $<\dots>_T$ is over the entire length $T$ of the time series considered. The denominator of the previous equation corresponds to the time averaged volatility $\sigma (\Delta t)$. For all data and for $\Delta t$ from 1 minute to 1 day, the distribution of returns is in agreement with a L\'evy distribution of index varying from $\alpha \approx 1.35$ to $\alpha \approx 1.8$ in the central part of the distribution, depending on the size of the region of empirical data used for the fit. In contrast to the previously suggested exponential truncation for the largest  returns, the distribution is found to scale like a power-law with an exponent $\mu  \equiv 1 + \alpha \approx 4$. This value is well outside the L\'evy stable regime, which requires $0< \alpha \le 2$. Hence, these study point towards a truncated L\'evy distribution, with a power-law truncation.

The previous scaling in time is observed for time scales $\Delta t$ of up to 4 days. For larger time scale, the data are consistent with a slow convergence towards a Gaussian distribution. This convergence is expected as the presence of a power-law cut-off implies that the distribution of returns is in the Gaussian stable regime, where the CLT applies. What is surprising is the existence of a `metastable' scaling regime observed in time scales as long as 4 days. Gopikrishnan {\it et al.} \cite{Gopi99} identified time dependencies as one of the potential sources of this scaling regime, by comparing the actual time series first to a randomly generated time series with the same distribution and second to the shuffled original time series. The two latters display a much faster convergence towards the Gaussian statistics, confirming their hypothesis.

One important conclusion of the previous empirical data is that the theory of truncated L\'evy distributions cannot reproduce empirical data as it stands, because in the current framework of truncated L\'evy distributions, the random variables are still independent, while it has been shown that time dependencies are a crucial ingredient of the scaling. Also, it does not explain the fluctuations of $\gamma$.
 
Plerou {\it et al.} \cite{Plerou00} considered the variance of individual price changes, $\omega^2_{\Delta t} \equiv <(\delta p_i)^2>$ for all transactions of the 1,000 largest US stocks for the 1994-1995 2-year period. The cumulative distribution of this variance displays a power-law scaling $P (\omega_{\Delta t}> x)\sim x^{-\gamma}$, with     $\gamma = 2.9\pm 0.1$. Using  detrended fluctuation analysis, they obtain a correlation function characterized by an exponent $\mu = 0.60\pm 0.01$, that is, weak correlations only, independent variables being characterized by $\mu = 1/2$.
 
Gopikrishnan {\it et al.} \cite{Gopi99} found that $\sigma (\Delta t) \sim (\Delta t)^{\delta}$. The exponent $\delta$ experiences a crossover from $\delta = 0.67\pm 0.03$ when $\Delta t< 20$ minute to $\delta = 0.51\pm 0.06$ when $\Delta t > 20$ minutes. This is in agreement with the fact that the autocorrelation function of the  returns is exponentially decreasing in a characteristic time $\tau_{ch}$ of 4  minutes. These results predict that for $\Delta t > 20$ minutes, the returns are essentially uncorrelated. Hence, important scaling properties of the time series originate from time dependencies, but the autocorrelation function of the  returns or the time averaged variance do not deliver the relevant information to study these dependencies. Higher order correlations need to be analyzed.
 
Since the time averaged volatility is a power-law, this invalidates the standard ARCH or GARCH processes, because they all predict an exponentially decreasing volatility. In order to explain the long range persistence of the correlations, one needs to include the entire history of the returns.

Scalas \cite{scalas} analyzed the statistics of the price difference between the closing price and the opening price of the next day, taking week-ends and holidays as overnight variations. He concentrated his analysis on Italian government bonds (BTP) futures for the 18 Sept 1991- 20 Feb 1997 period. As for other time scales variations, he did not observe short-range nor long-range correlations in the price changes. In fact he was able to reproduce similar results with a numerical simulation of a truncated trinomial random walk.

A general framework for treating superdiffusive systems is provided by the nonlinear Fokker-Planck equation, which is associated with an underlying Ito-Langevin process. This in turn   has a very interesting connection to the nonextensive entropy proposed by Tsallis : the  nonlinear Fokker-Planck equation is solved by time-dependent distributions which maximize the Tsallis entropy. This unexpected connection between thermostatistics and anomalous diffusion gives an entirely new way, much beyond Bachelier \cite{Bach}   like approach, to study the dynamics of financial market as if there are  anomalously diffusing systems.

Whence the intra-day price changes in the S\&P500 stock index have been studied within this framework by direct analysis and by simulation in refs. \cite{michaeljohnso,MAKIPRE,borland}. The power-law tails of the distributions, and the index's anomalously diffusing dynamics, are very accurately described. Results show good agreement between market data, Fokker-Planck dynamics, and simulation. Thus the combination of the Tsallis non-extensive entropy and the nonlinear Fokker-Planck equation unites in a very natural way the power-law tails of the distributions and their superdiffusive dynamics.  In our case the shape and tails of partial distribution functions (PDF) for a financial signal, i.e. the S\&P500 and the turbulent nature of the markets were linked through Beck  model, originally proposed to describe the intermittent behavior of turbulent flows. 
Looking at small and large time windows, both for small and large log-returns. The market volatility (of normalized log-returns) distributions was  well fitted with a $\chi^2$-distribution. The transition between the small time scale  model of nonextensive, intermittent process and the large scale  Gaussian extensive homogeneous fluctuation picture was found to be at $ca.$ a 200 day time lag. The intermittency exponent ($\kappa$) in the framework of the Kolmogorov log-normal model was found to be related to the scaling exponent of the PDF moments, -thereby giving weight to the  model. The large value of $\kappa$ points to a large number of cascades in the turbulent process. The first  Kramers-Moyal coefficient in the  Fokker-Planck equation is almost equal to zero, indicating ''no restoring force''. A comparison is made between normalized log-returns and mere price increments.

\begin{figure}[b]
\includegraphics[width=.75\textwidth]{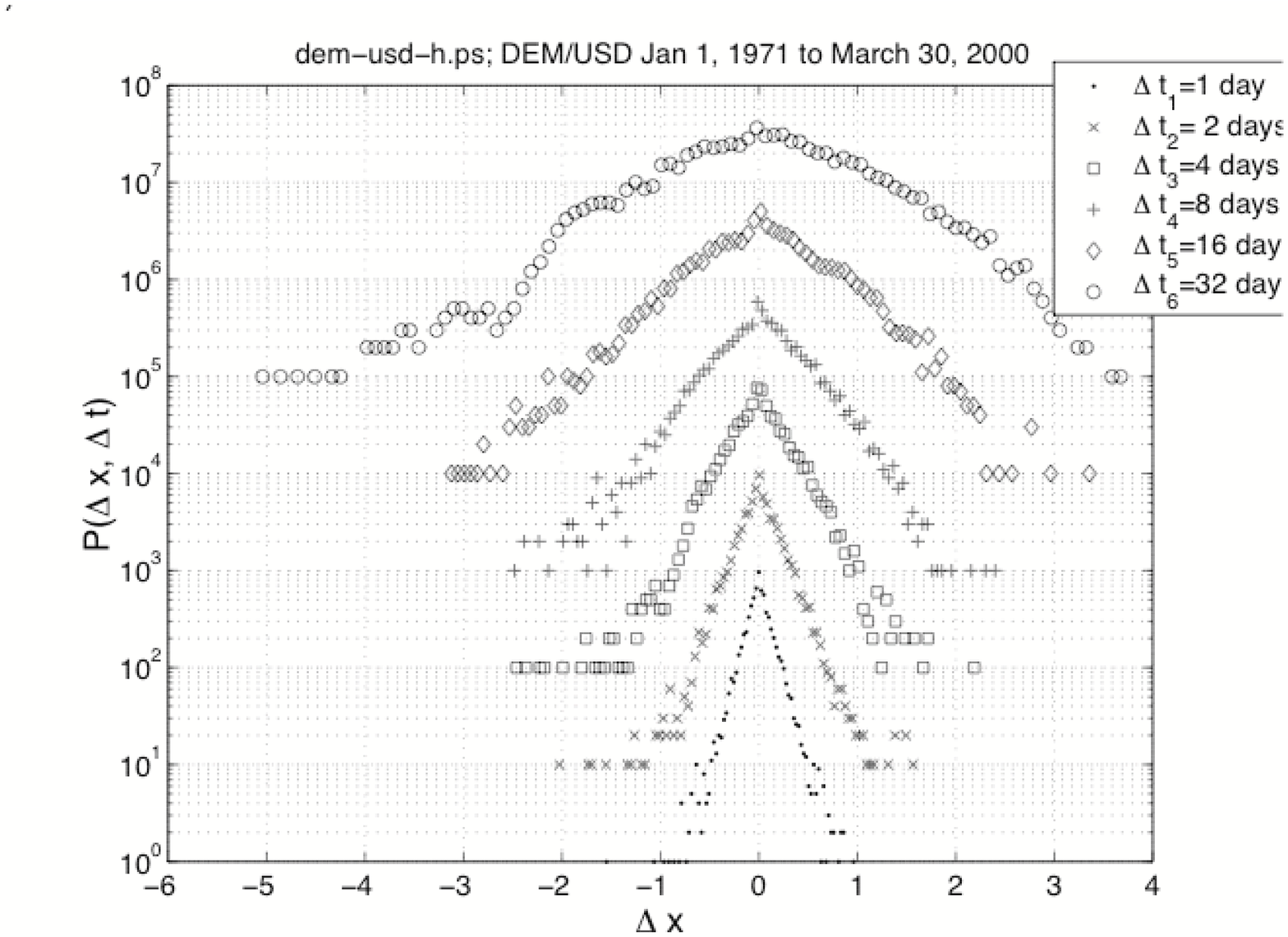}

\vskip 1.0cm
\caption{Probability distribution function $P(\Delta x, \Delta t)$ of normalized increments $\Delta x$ of daily closing price value signal of $DEM-USD$ exchange rate between Jan. 01, 1971 and March 31, 2000, for  different time lags: $\Delta t=1,2,4,8,16,32$~days. The normalization is with respect to the
width of the PDF for $\Delta t= 32$~days. The PDF symbols for a given $\Delta t$ are displaced by a factor 10 with respect to the previous $\Delta t$ ; the curve for $\Delta t=1$~day is unmoved. See the
tendency toward a gaussian for increasing $\Delta t$
\label{fig1}}
\end{figure}
\begin{figure}[b]
\includegraphics[width=.75\textwidth]{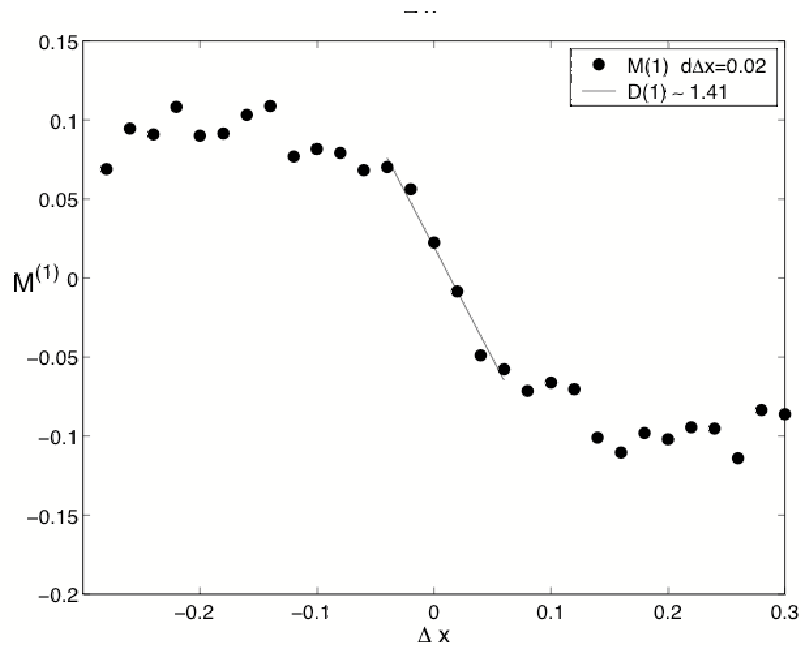}

\vskip 1.0cm
\caption{Kramers-Moyal drift  coefficient    $M^{(1)}$  as a function of normalized increments
  $\Delta x$ of daily closing price value  of $DEM-USD$ exchange rate between Jan. 01, 1971 and March 31, 2000, with a best linear fit for the central part of the data corresponding to a drift coefficient $ D^{(1)} =  -1.41$
\label{fig2}}
\end{figure}

\begin{figure}[b]
\includegraphics[width=.75\textwidth]{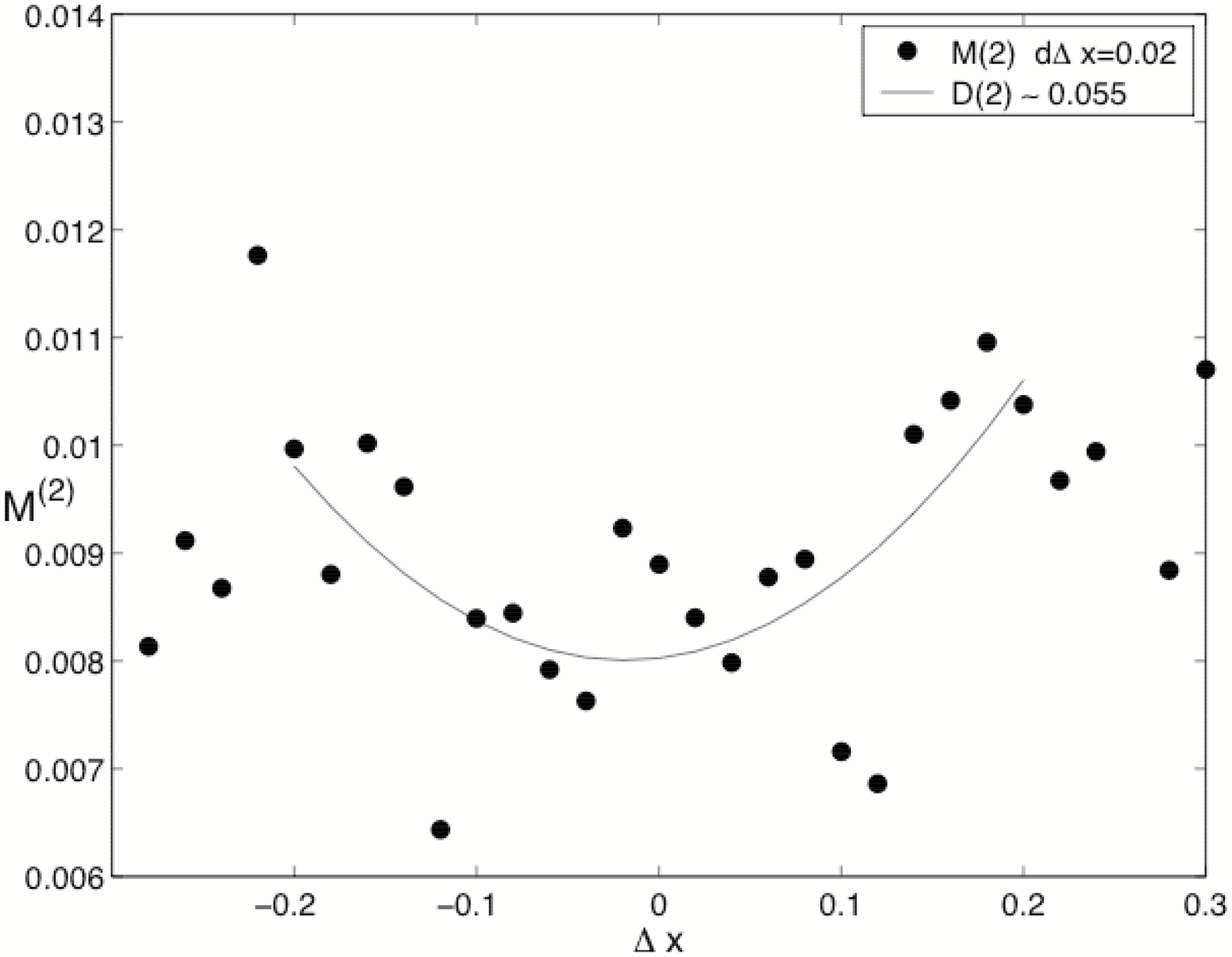}

\vskip 1.0cm
\caption{Kramers-Moyal diffusion coefficient  $M^{(2)}$ as a function of normalized increments
  $\Delta x$ of daily closing price value  of $DEM-USD$ exchange rate
between Jan. 01, 1971 and March 31, 2000, with a best parabolic fit for
the central part of the data
corresponding to a diffusion coefficient $ D^{(2)} =  0.055$
\label{fig3}}
\end{figure}

The case of foreign exchange markets might have not been studied to my 
knowledge, and might be reported here for the first time 
(Fig. \ref{fig1}).  Consider   
the time series of the normalized log returns 
$Z(t,\Delta t)=(\tilde y(t) - <\tilde y>_{\Delta t})/\sigma_{\Delta t}$  for different (selected) values of the time lag  $\Delta t=1,2,4,8,10,32$~days.  Let $\tau=log_2(32/\Delta t)$, 
\begin{equation} 
\frac{d}{d\tau}p(Z,\tau )=\left[-\frac{\partial }{\partial Z} D^{(1)}(Z,\tau )+\frac{\partial }{\partial Z^{2}} D^{(2)}(Z,\tau )\right]p(Z,\tau) \label{efp} 
\end{equation} 
in terms of a drift $D^{(1)}$($Z$,$\tau $) and a diffusion coefficient $D^{(2)}$($Z$,$\tau $) (thus values of $\tau $ represent $ \Delta t_{i}$, $i=1,...$).

The coefficient functional dependence can be estimated directly from the moments $M^{(k)}$ (known as  Kramers-Moyal coefficients) of the conditional probability distributions:   
\begin{equation} 
M^{(k)}=\frac{1}{\Delta \tau }\int dZ^{^{\prime }} (Z^{^{\prime}}-Z)^{k}p(Z^{^{\prime }},\tau +\Delta \tau |Z,\tau ) 
\end{equation}

\begin{equation} 
D^{(k)}(Z,\tau )=\frac{1}{k!}\mbox{lim} M^{(k)} 
\end{equation} 
for $\Delta \tau \rightarrow 0$. According to Fig. \ref{fig2}  
the drift coefficient  $D^{(1)}\approx0$ and the diffusion coefficient 
$D^{(2)}$ (Fig. \ref{fig3}) is not so well represented by a parabola 
as for the $S\&P$.  A greater deviation from regular curves is also seen 
if one examines specific stock evolutions.

It may be   recalled that the observed quadratic dependence of the diffusion 
term $D^{(2)}$ is essential for the logarithmic scaling of the
intermittency parameter in studies on turbulence.

\subsection{Crashes}
 
Johansen and Sornette \cite{JoSo98} stressed that the largest  crashes of the $XX-th$ century appear as outliers, not belonging to the distribution of \textit{normal}  returns. The claim is easily observable for the three major crashes (downturns of more than 22\%), October 1987, World War I and Wall Street 1929, in order of magnitude. It is also supported by the fact that in a simulation of a million-year trading using a GARCH(1,1) model, with a reset every century, never did 3 crashes occur within the same century. Another important property that points towards a different dynamical process for crashes is that, if anomalously large returns are for instance defined as a change of 10\% or more of an index over a short time interval (a couple of days at most), one observes only  crashes (downturns) and no upsurge. This contrasts with usual  returns which have been observed to be symmetric for up and down moves.

In agreement with  crashes being not part of the usual distribution of  returns, Feigenbaum and Freund \cite{JFeFr96} and Sornette, 
Johansen and Bouchaud \cite{SoJoBo96} have independently proposed a picture of  crashes as critical points in a hierarchical system with discrete scaling. When a system is at a critical point, it becomes `scale invariant', meaning that it becomes statistically similar to itself when considered at different magnifications. The signature of self-similarity is power-laws, which are scale free laws. For systems with a hierarchical structure, the system is self-similar only under specific scale changes. The signature of this property is that the exponent of the previous power-laws has a non-zero imaginary part. Specifically, if $y( t)$ is the price at time $t$, it is supposed that

\begin{equation} 
y (t) = A + B (t_c - t)^{\mu } [1 + C \cos [\omega \log (t_c-t) + \phi]] 
\label{eq:log-periodic oscillations}     
\end{equation} 
in the regime preceeding a crash. 
The previous equation stipulates that 
$t_c$ is a critical time where the prices will experience of phase transition. The signature of this transition is a variation of the price as  a power-law of exponent $\mu$. But because of the hierarchical structure in the system, the divergence is `decorated' by log-periodic oscillations. The phase $\phi$ defines the chosen unit of time scale. It has appeared that the value of $\mu$ is not robust in the non-linear fits. 

In a latter study, the S\&P500 and the Dow Jones 1987 crash were examined \cite{VBMA98} after subtracting an exponential background due to the time evolution of these indices before a fitting with Eq. (\ref{eq:log-periodic oscillations}). Next it has been proposed to consider a logarithmic divergence, corresponding to the $\mu=0$ limit, \cite{how} rather than a power law,  i.e.  
\begin{equation} 
y(t) = A + B \ln{\left( {t_c-t \over t_c} \right)} \left[ 1 + C sin \left(\omega \ln{\left ( {t_c-t \over t_c} \right)} + \phi \right) \right]  \hskip 0.4cm for \hskip 0.2cm t < t_c 
\end{equation}

In so doing the analysis of (closing value)  stock market index like the Dow Jones Industrial average ($DJIA$), the S\&P500 \cite{how,vizlogp} and $DAX$ \cite{drzw} leads to observe the precursor of so-called  crashes. This was shown on the Oct. 1987 and Oct. 1997 cases, as it has been reported in the financial press in due time \cite{dup1,dup2}. The prediction of the crash date was made as early as July, in the 1997 case. It should be pointed out that we do not expect any real divergence in the stock market indices, but to give upper bounds are surely of interest.
However, the reliability of the method has been questioned by Laloux {\it et al.} \cite{laloux2} on the  grounds that all empirical fits require a large number of parameters, that the statistical data for  crashes is obviously very restricted and that the method predicts crashes that never happened. This can be further debated at other places.

Yet the existence of log-periodic corrections to scaling implies the existence of a hierarchy of time scales $\delta t_n \equiv t_c - t_n$ determined by half-periods of the cosine in Eq. (\ref{eq:log-periodic oscillations}). These time scales are not expected to be universal, varying from a market to the other, but the ratio $\lambda = \delta t_{n+1} / \delta t_n$ could be robust with respect to changes in what is traded and when it is traded, with $\lambda \approx 2.2-2.7$, see Sornette \cite{DSphysrep}.
 
Sornette, Johansen and Bouchaud \cite{SoJoBo96} identify the underlying hierarchical structure as having the currency and trading blocks at its highest level (Euro, Dollar, Yen,...), countries at the level below, major banks and institutions within a country at the level below, the various departments of these institutions at the level below, and so on listed Several potential sources can be listed \cite{VBMA98}  for discrete scaling: hierarchical structure of investors (see Table \ref{table:investor types}), investors with specific time horizon or periodic events like market closure and opening, quaterly publication of reports, ...

\begin{table}[h]
\caption{Different categories of investors and their relative weight on the London Stock Market in 1993 after David
\cite{david94}}\label{table:investor types}
\begin{small}\sffamily
\begin{tabularx} {243pt}{|c|c|c|}
\hline

\textbf{Rank} & \textbf{Investor type} & \textbf{Weight}\\

\hline

1 & Pension Funds & 34.2\\
2 & Individuals & 17.7\\
3 & Insurance companies & 17.3\\
4 & Foreign & 16.3\\
5 & Unit Trusts & 6.6\\
6 & Others & 2.3\\
7 & Other personal sector & 1.6\\
8 & Industrial and commercial companies & 1.5\\
9 & Public Sector & 1.3\\
10 & Other Financial Institutions & 0.6\\
11 & Banks & 0.6\\

\hline
\end{tabularx}
\end{small}
\end{table}
 
Feigenbaum and Freund \cite{JFeFr96} had reported log-periodic oscillations for 1986-1987 weekly data for the S\&P500 (October 19, 1987 crash), 1962 monthly data for the S\&P500 (NY crash of 1962), 1929 monthly data for the Dow-Jones (NY 1929 crash) and 1990 scanned data for the NIKKEI (Tokyo 1990 crash). 
Sornette, Johansen and Bouchaud\cite{SoJoBo96}  had also reported log-periodic oscillations for July 1985 to the end of 1987 daily data for the S\&P500. Since then, this type of oscillations have been detected before many financial indices downturns, see refs. \cite{how,vizlogp} and ref. \cite{DSphysrep} for a review.
 
Johansen and Sornette \cite{JoSo98} have also analyzed the symmetric phenomenon with respect to the time, that is, a power-law decrease of the price after culmination at a maximum value, the power-law being decorated by log-periodic oscillations. Interestingly, they do not obtain empirical evidence of a maximum with log-periodic oscillations before \emph{and} after the maximum. The authors advocate as main reason that when the market accelerates to reach a log-periodic regime, it very often ended in a  crash, with an associated symmetry breaking. According to this argument, the symmetrical phenomenon can only be detected whenever a crash did not happen before. They found empirical evidence of `antibubbles' for the NIKKEI, for the 31 Dec. 1989 till 31 Dec. 1998 period and for gold futures after 1980. It seems that $\lambda$ for decreasing markets could be higher (around $3.4-3.6$) than for increasing markets.

\subsection{Crash models}
 
Eliezer and Kogan \cite{ElKo98} have considered the possibility of the existence of a crashing phase, with different dynamics from the normal or quiescent phase of the market. Their extension is based on a model with two types of traders, but different from the previous noise and rational traders. The distinction comes from the way the agents place their orders. Agents that place a limit order are visible to every one on the screen, while agents placing market orders are not. Hence, it is impossible to have an annihilation event between agents placing market orders, as they are `invisible' to each others. This distinction between market order agents and limit order agents was emphasized by Cohen {\it et al.} \cite{cohen81} as a possible source for widening the spread between ask and bid prices.  

\subsection{The Maslov model}
 
Maslov \cite{Maslov} proposed a model similar to the crash model of ref. \cite{ElKo98}. He assumes that each agent can either trade a stock at the market price or place a limit order to sell or buy. At each time step, a new trader appears and tries to make a transaction, buying or selling being equally likely to occur. With probability $q_{lo}$, he places a limit order, otherwise, he trades at the best limit order available, the lowest ask for a buyer  or the highest bid for a seller. The price of the last deal is taken as the market price.  The price of a new limit order is equal to the market price offset by a random number $\Delta$ chosen from a probability distribution, to avoid overlapping buy and sell orders. Agents making a deal are removed from the system. In contrast to the previous models, the number of agents is not fixed and the agents are not able to diffuse. Once they have chosen a price, they can only wait until someone matches their order.

Numerically, it is obtained that the model displays volatility clustering, a power-law decrease of the autocorrelation function of absolute price change with an exponent $\gamma = 1/2$ (0.3 on real markets), a non-trivial Hurst exponent $H = 1/4$ (0.6-0.7 on real markets), no correlation on the signs of the price changes. Most importantly, the price increment distribution is non-Gaussian, with a crossover between two power-laws in the wings, from an exponent $1 + \alpha_1$ to $1+ \alpha_2$ for larger changes, with $\alpha_1 = 0.6\pm 0.10$ and $\alpha_2 =3.0 \pm 0.2$. In summary, the model is very promising because it has many qualitative interesting features, but unfortunately a chronic lack of quantitative agreement with real data.

Within a mean-field approximation of the model, Slanina \cite{Slanina} has shown that the stationary distribution for the price changes has have power-law tails with an exponent $1 + \alpha = 2$. This result disagrees with numerical simulations of the model, both quantitatively and qualitatively, for reasons that remain unknown.

\subsection{The sandpile model}

It can be conjectured that  stock markets are hierarchical objects where each level has a different weight and a different characteristics time scale (the horizons of the investors). The hierarchical lattice might be a fractal tree \cite{vizlogp} with loops. The geometry might control the type of criticality.  This led to consider the type of avalanches which occur in a tumbling sandpile as an analogy  with the spikes in the index variation within a financial bubble grow.
 
The Bak, Tang and Wiesenfeld (BTW) \cite{btw} in its sandpile version \cite{bak2} was studied on a Sierpinski gasket   \cite{gasket}. It has been shown that the avalanche dynamics is characterized by a power law with a complex scaling exponent $\tau + i \omega$ with $\omega = \frac{2\pi}{\ln{2}}$. This was understood as the result of the underlying Discrete Scale Invariance (DSI) of the gasket, i.e. the lattice is translation invariant on a log-scale \cite{DSphysrep}.  Such a  study of the BTW  model was extended to studies in which were varying both the fractal dimension $D_f$ as well as the connectivity of the lattice. In so doing  connectivity-based corrections to power law scaling appeared. For most avalanche distributions $P(s)$ $\sim$ $s^{-\tau} $, expressing the scale invariance of the dynamics, we have checked the power-law exponent ($\tau$) as a function of the fractal dimension of RSC lattices and have found that $\tau$ seems to be dependent of the real part of the lattice (carpet) fractal dimension $\Re \{ D_f \}$. Notice that for $\Re \{ D_f \} \rightarrow 2$, $\tau = 1.25 \pm 0.03$.

We have observed significant deviations of $\tau$ from 1.25, i.e. the $d=2$ value. It seems that the real part of the fractal dimension of the dissipative system is not the single parameter controlling the value of $\tau$.    These oscillations (peaks) can be thought to originate from the DSI of the RSC lattice as in \cite{gasket}, and to mimic those discussed in the main text for financial indices.  We emphasize that the connectivity of the lattice is one of the most relevant parameters. Notice that such log-periodic oscillations and linearly substructured peaks are observed in the time distribution of avalanche durations $P(t)$ as well.

\subsection{Percolation models}

Percolation theory was pioneered in the context of gelation, and later introduced in the mathematical literature by Broadbent and Hammersley \cite{DSAA91,Sahi94} with the problem of fluid propagation in porous media. Since then, it has been successfully applied in many different area of research : e.g., earthquakes, electrical properties and  immunology. Its application to financial market leads to one major conclusion: herding is a likely explanation for the fat tails \cite{ContBouchaud1997}. Percolation  models assume {\it a contrario} to usual financial theories that the outcomes of decisions of individual agents may not be represented as independent random variables. Such an assumption would ignore an essential ingredient of market organization, namely the {\it interaction} and {\it  communication} among agents. In real markets, agents may form groups of various sizes which then may share information and act in coordination. In the context of a financial market, groups of traders may align their decisions and act in unison to buy or sell; a different interpretation of a `group' may be an investment fund corresponding to the wealth of several investors but managed by a single fund manager.

To capture the effect of the interactions between agents, it is assumed that market participants form groups or ``clusters" through a random matching  process but that no trading takes places inside a given group: instead, members of a given group adopt a common market strategy (for example, they decide to buy or sell or not to trade) and different groups may trade with each other through a centralized market process. In the context of a financial market, clusters may represent for example a  group of investors participating in a mutual fund. 
 
\subsection{The Cont-Bouchaud model}

The Cont-Bouchaud  model is an application of bond percolation on a random graph, an approach first suggested by Kirman
\cite{kirman} in the economics literature. Each site of the graph is supposed to be an agent, which is able to make a transaction of one unit on a market. Two agents are connected with each other with probability $p = c/N$. At any time, an agent is either buying with probability $a$, selling with probability $a$ or doing nothing with probability $1-2a$, with 
$a\in (0,1/2)$.

The exogenous parameter $a$ controls the transaction rate, or the time scale considered; $a$ close to 0 means short time horizon of the order of a minute on financial markets, with only a few transactions per time steps. The number of traders which are active during one time interval increases with the length of this time interval. All agents belonging to the same cluster are making the same decision at the same time. The cluster distribution represents the network of social interactions, which induces agents to act cooperatively. This network is supposed to model the phenomenon of herding. The aggregate excess demand for an asset at time $t$ is the sum of the decision of all agents,

\begin{equation}
  D(t) =  \sum _{i=1}^{N} \phi_{i}(t),
\end{equation}
if the demand of agent $i$ is $\phi_{i} \in \{-1,0,+1\}$,  $\phi_{i}=-1 $ representing a sell order.     $D (t)$ is not directly accessible, so that it has to be related to the returns through some  $R = F (D).$

There is unfortunately no definite consensus about the analytic form of $F$. This will prevent a convincing quantitative comparison between the model and empirical data. Nevertheless, for purpose of illustration, we will assume that the    price change during a time interval is proportional to the sum of the demand and sell orders from all the clusters which are active during this time interval. That is, we consider $F (D) \sim D$, unless specified otherwise. 
For $c=1$ the probability density for the cluster size distribution decreases asymptotically as a power law

\begin{equation}
n_s  {\sim}_{s\to\infty} \frac{A}{s^{5/2}}
\end{equation} 
while for $0 < 1-c << 1$, the cluster size distribution is cut off by an exponential tail,
\begin{eqnarray}
n_s  {\sim}_{s\to\infty} \frac{A}{s^{5/2}} \exp 
\left(-\frac{(c-1)s}{s_0} \right) \label{tail}
\end{eqnarray}
For $c$=1, the distribution has an infinite variance while for $c < 1$ the variance becomes finite because of the exponential tail. In this case  the average size of a coalition is of order $1/(1-c)$ and the average number of clusters is of order $N(1- c/2)$. Setting the coordination parameter $c$ close to 1 means that each agent tends to establish a link with one other agent. In the limit $N\to\infty$, the number $\nu_i$ of neighbors of a given agent $i$ is a Poisson random variable with parameter $c$,
\begin{eqnarray}
P(\nu_i = \nu) &=& e^{-c}\frac{c^{\nu}}{{\nu}!}
\end{eqnarray}

The previous results are based in the assumption that either exactly  one cluster of traders is making a transaction at each time step or none. If the time interval of observation is increased enough to allow each cluster to be considered once during a time step, numerous clusters could decide to buy or sell, depending on the value of $a$. In this case, the probability distribution of the returns will be different from the cluster size distribution. By increasing $a$ from close to 0 to $a$ close to 1/2,  Stauffer and Penna (1998) have shown that the probability distribution of the returns changes from a exponentially truncated power-law distribution to a Gaussian distribution. At intermediate value, a continuous distribution with a smooth peak and power-law in the tails only is obtained, in agreement with a Levy distribution. This situation is similar to a change in time scale. At short time scales, like every minute, there is either an order that is placed, or none, while for longer time scales, the observed variations on financial markets are the average result of thousands of orders. Increasing $a$ from 0 to $1/2$ is similar to changing the time scale from short-time scales to large-time scale. Gopikrishnan {\it et al.} \cite{Gopi99} have observed empirically this convergence towards a Gaussian distribution by changing the time scale for financial market indices.  
Recall that Ausloos and Ivanova \cite{MAKIPRE} discussed the fat tails in another way, arguing that they   are caused by  some ''dynamical process'' through a hierarchical cascade of short and long-range volatility correlations. Unfortunately, there are no correlations in the  Cont-Bouchaud model  to compare with this result.

The Cont-Bouchaud model predicts a power-law distribution for the size of associations, with an exponential cut-off. When $c$ reaches one, a finite fraction of the market shares  simultaneously the same opinion and this leads to a crash. Unfortunately, this crash lacks any precursor pattern because of the lack of time correlations, amongst other possible shortcomings. From the expected relation $R = F (D)$, it leads similarly to power-law tails for the distribution of  returns. However, the exact value of the exponent of this power-law is still a matter of debate because of the lack of consensus upon $F (D)$. We report in Table \ref{table:cont-bouchaud} a summary of the agreement of the model and its limitations.
\begin{table}[h]
\caption{Summary of the type of agreement between empirical data and  Cont-Bouchaud  model}
\label{table:cont-bouchaud}
\begin{small}\sffamily
\begin{tabularx} {113pt}{|c|c|}
\hline

\textbf{Property} & \textbf{Agreement}\\

\hline
Fat tails & Qualitative \\
Crossover &  No \\
Symmetric & Yes \\
Clustering & No \\
Crashes &  Yes \\
Precursors & No \\

\hline
\end{tabularx}
\end{small}
\end{table}

Eguil\'uz and Zimmermann \cite{EgZi00} introduced a dynamical version of the Cont-Bouchaud model. At each time step, an agent is selected at random. With probability $1-2a$, she selects another agent and a link between those two agents is created. Otherwise, this agent triggers her cluster to make a transaction, buying with probability $a$, selling with probability $a$. 
Eguil\'uz and Zimmermann associated the creation of a link with the exchange of information between agents, while when the agents make a transaction, they make their information public, that is, this information is lost. 
D'Hulst and Rodgers \cite{DHuRo00} have shown that this model is equivalent to the Cont-Bouchaud  model, except that with the new interpretation, the probability that a cluster of size $s$ makes transaction is $sn_s$ rather than $n_s$ as in the Cont-Bouchaud model. Extension of the   model   allowing for the cluster distribution not to change can be envisaged.

Finally, we should mention that numerical simulations show that it is possible to observe power-laws with higher than expected effective exponent. In other words, size effects can modify the exponent. As all empirical data are strongly affected by size effects, looking after a  model that reproduces the exact value of the exponent seems less important than trying to justify why the exponent should have a given value.

\subsection{Crash precursor patterns}

In the  Cont-Bouchaud model for $p\ge p_c$, there is a non-zero probability that a finite fraction of agents decide to buy or sell. These large cooperative events generate anomalous wings in the tails of the distribution, that is, it appears as outliers of the distribution. The same pattern has been observed by 
Johansen and Sornette   for crashes on financial markets, which means that the decision from buying or selling of the percolating cluster can be compared to a crash. These crashes however lack any precursor patterns, which have been empirically observed for real  crashes, as originally proposed by Sornette, Johansen and Bouchaud.
 
As explained by Stauffer and Sornette \cite{StauSor} log-periodic oscillations, a  crash precursor signature, can be produced with biased diffusion. The sites of a large lattice are randomly initialised as being accessible with probability $p$, or forbidden, with probability $1-p$. A random-walker diffuses on the accessible sites only. With probability $B$, the random-walk moves in one fixed direction, while with probability $1-B$, it moves to one randomly selected nearest neighbour. In both cases, the move is allowed only if the neighbour is accessible. Writing the time variation of the mean-square displacement from a starting point as $<r^2> \sim t^k$, $k$ changes smoothly as a function of the logarithm of the time. That is, $k (t)$ approaches unity while oscillating according to $\sin (\lambda \ln t)$. Behind these log-periodic oscillations is the fact that the random-walker gets trapped in clusters of forbidden sites. The trapping time depends exponentially on the length of the traps, that are multiple of the lattice mesh size. The resulting intermittent random-walk is thus punctuated by the successive encounters of larger and larger clusters of trapping sites. This result is much more effective in reduced dimensionality, or  equivalently when $B \to 1$.

We have to stress that, even if this  model is very closely related to  percolation, there exists no explanation on how to relate the distance traveled by a diffusing particle to transactions on financial markets. Connection to the sandpile model is still to be worked out. Hence, this model is not an extension of the Cont-Bouchaud model of financial markets, but rather a hint of a possible direction for on how implement log-periodic oscillations. 
   
Notice that within the framework of the Langevin equation proposed by Bouchaud and Cont \cite{BouCon98}, a crash occurs after an improbable succession of unfavorable events, because they are initiated by the noise term. Hence, no precursor pattern can be identified within the original equation. Bouchaud and Cont extended their model and proposed the following mechanism to explain log-periodic oscillations. Every time an anomalously negative value $u$ close to $u^{*}$ is reached, the market becomes more `nervous'. This is similar to saying that the width $W$ of the distribution describing the noise term $\eta$ increases to $W + \delta W$. Therefore, the time $\Delta t$ between the next anomalously large negative value will be shorten as large fluctuations become more likely. The  model predicts

\begin{equation}
\Delta t_{n+1} = \Delta t_n S^{- \delta W / W}
\end{equation} 
where, to linear order in $\delta W$, $S$ is some constant. This leads to a roughly log-periodic behaviour, with the time difference between two events being a geometric series. The previous scenario predicts that a  crash is not related to a critical point. That is, there is a crash because $u$ reaches $u^{*}$, not because $\Delta t \to 0$.

\section{The Lux-Marchesi model}
 
Lux and Marchesi \cite{LuxMarchesi99} have introduced a model of financial market where the agents are divided in two groups, fundamentalists and noise traders. Fundamentalists expect the price to follow the so-called fundamental value of the asset ($p_f$), which is the discounted sum of expected future earnings (for example, dividend payments). A fundamentalist strategy is to buy (sell) when the actual price is believed to be below (above) $p_f$. The noise traders attempt to identify price trends and patterns and also consider the behaviour of other traders as a source of information. The noise traders are also considered as optimistic or pessimistic. The former tends to buy because they thing the market will be raising, while the latter bet on a decreasing market and rush out of the market to avoid losses.

The dynamics of the model are based on the movements of agents from one group to the other. A switch from one group to the other happens with a certain exponential probability $\nu e^{U (t)} \Delta t$, varying with time. $\nu$ is a parameter for the frequency of revaluation of opinions or strategies by the agents.  $U (t)$ is a forcing term covering the factors that are relevant for a change of behaviour, and it depends on the type of agent considered. Noise traders use the price trend in the last time steps and the difference between the number of optimistic and pessimistic traders to calculate $U (t)$ corresponding to transitions between the group of optimists and pessimists. For example, observation of a positive trend and more optimists than pessimists is an indication of a continuation of the rising market. This would induce some pessimistic agents to become optimistic. The other type of transitions is between fundamentalists and noise traders. The $U (t)$ function corresponding to such transitions is a function of the difference in profits made by agents in each group. An optimistic trader profit consists in short-term capital gain (loss) due to price increase (decrease), while a pessimistic trader gain is given by the difference between the average profit rate of alternative investments (assumed to be constant) minus the price change of the asset they sell. A fundamentalist profit is made off the difference between the fundamental price and the actual price, that is, $|p-p_f|$ is associated to an arbitrage opportunity. In practice, Lux and Marchesi (1999) multiply this arbitrage profit by a factor $<1$ to take into account the time it takes some time for the actual price to reverse to its fundamental value.

To complete the model, it remains to specify how the price and its fundamental value are updated. Price changes are assumed to be proportional to the aggregate excess demand. Optimistic noise traders are buying, pessimistic noise traders are selling and fundamentalists are all buying or all selling depending if $p_f - p$ is positive or negative, respectively. Finally, relative changes in the fundamental price are Gaussian random variable, $\ln p_{f,t} - \ln p_{f,t-1} = \epsilon_t$, where $\epsilon_t$ is a normally distributed random variable, with mean zero and time-invariant variance $\sigma^2_{\epsilon}$.
 
Lux and Marchesi claimed that a theoretical analysis of their model shows that the stationary states of the dynamics are characterized by a price which is on average equal to its fundamental value. This is supported by numerical simulations of the  model, where it can be seen that the price tracks the variation of its fundamental value. But comparing price returns, by construction, the fundamental returns are Gaussian, while price returns are not. The distribution of price  returns display fat tails that can be characterized by a power-law of exponent $\tau = 1 +\alpha$. Numerically, Lux and Marchesi obtained $\alpha = 2.64$ when sampling at every time step. Increasing the time lag, a convergence towards a Gaussian is observed. They measured a Hurst exponent $H = 0.85$ for the price returns, showing strong persistence in the volatility. The exact value of the exponent is close to empirical data.

The behaviour of the model comes from an alternation between quiet and turbulent periods. The switch between the different types of periods are generated by the movements of the agents. Turbulent periods are characterized by a large number of noise traders. There exists a critical value $N_c$ for the number of traders, such as when there are more than $N_c$ traders, the system looses stability. However, the ensuing destabilization is only temporary as the presence of fundamentalists and the possibility to become a fundamentalist ensure that the market always stabilizes again. These temporary destabilizations are known as on-off intermittency in physics.

\subsection{The spin models}

There are necessarily crash and financial market  aspects which resemble phase transitions; this has  led into producing a realm of spin models.  The superspin  model proposed by 
Chowdhury and Stauffer \cite{ChowStau99} is related to the Cont-Bouchaud model presented here above. The cluster of connected agents are replaced by a superspin of size $S$. $|S_i|$ is the number of agents associated with the superspin $i$. The value of $|S_i|$ is chosen initially from a probability distribution $P (|S_i|) \sim |S_i|^{-(1+\alpha)}$. A superspin can be in three different states, $+|S_i|$, 0 or $-|S_i|$. Associated to each state is a `disagreement function' $E_i = - S_i (H_i + h_i)$, where $H_i = J \sum_{j\not = i} S_j$ is a local field and $h_i$ an individual bias. $E_i$ represents the disagreement between the actual value of a superspin and two other factors. One of these factors, $H_i$, is an average over the decision of the other groups of agents. The other factor, $h_i$, is a personal bias in the character of a group of agents, with groups of optimists for $h_i >0$, and pessimists for $h_i < 0$. Groups with $h_i = 0$ are pure noise traders. At each time step, every superspin $i$ chooses one of its three possible states, $+|S_i|$ with probability $a$ (buying) , $-|S_i|$ with probability $a$ or 0 with probability $1-2a$ (doing nothing). A superspin is allowed to change from its present state to the new chosen state with probability $e^{-\Delta E_i /k_b T}$, where $\Delta E_i$ is the expected change in its disagreement function. The magnetization $M$ corresponds to the aggregate excess demand, $D$.

If a linear relation $F (D)$ is assumed between the returns and $D$, the previous model is characterized by a probability distribution for the prices with power-law tails with the same exponent $1 + \alpha$, as the distribution of spins. In contrast to the Cont-Bouchaud model, this result stays true for all values of $a$, that is, there is no convergence towards a Gaussian for any value of $a$, contrary to what is observed on financial markets. Using $h_i$, it is possible to divide the population in noise traders for $h_i \ll H_i$ and fundamentalists for $h_i \gg H_i$. It is also possible to introduce a dynamics in $h_i$ to reflect the different opinions in different states of the market. No complete study of these properties has been done to date.

\section*{Acknowledgments}

The author thanks Ren\'e D'Hulst for an important contribution to this work.
MA would like to thank the many coworkers who helped  to make this review possible, i.e., in the last few years, Ph. Bronlet and K. Ivanova. Part of this work has been supported through the Minister of Education under contract ARC (94-99/174) and (02/07-293) of ULg.

\def\bibindent{6mm}

\include{fileFig}
\include{tabelle_articolo}
\end{document}